\documentclass[10pt,conference, anonymous]{IEEEtran}
\IEEEoverridecommandlockouts
\usepackage[noend]{algpseudocode}
\usepackage{amsmath,amssymb,amsfonts}
\usepackage[labelformat=simple]{subcaption}

\usepackage[hidelinks]{hyperref}
\usepackage[utf8]{inputenc}
\usepackage{dblfloatfix}
\usepackage{algorithm}
\usepackage{tcolorbox}
\usepackage{graphicx}
\usepackage{booktabs}
\usepackage{textcomp}
\usepackage{graphicx}
\usepackage{cleveref}
\usepackage{multirow}
\usepackage{textcomp}
\usepackage{balance}
\usepackage{comment}
\usepackage{siunitx}
\usepackage{diagbox}
\usepackage{xcolor}
\usepackage{pifont}
\usepackage{bm}

\newcommand{\xmark}{\ding{55}}

\usepackage[
backend=biber
,style=ieee
,minbibnames=1
,maxbibnames=99
,maxcitenames=1
,mincitenames=1
,eprint=false
,doi=false
,isbn=false
,url=false
]{biblatex}
\bibliography{ref} 
\AtBeginBibliography{\footnotesize}

\begin{document}

\title{ Bug or not Bug? Analysing the Reasons Behind Metamorphic Relation Violations}

\author{\IEEEauthorblockN{
	Alejandra Duque-Torres\IEEEauthorrefmark{2},
	Dietmar Pfahl\IEEEauthorrefmark{2}, 
    Claus Klammer\IEEEauthorrefmark{3}
    and
    Stefan Fischer\IEEEauthorrefmark{3}}
			
	\IEEEauthorblockA{%
	\IEEEauthorrefmark{2}\textit{Institute of Computer Science  }, \textit{University of Tartu}, Tartu, Estonia \\
	E-mail: \{duquet, dietmar.pfahl\}@ut.ee}	
	\IEEEauthorblockA{
	\IEEEauthorrefmark{3}\textit{Software Competence Center Hagenberg (SCCH) GmbH}, Hagenberg, Austria \\
	E-mail:  \{claus.klammer, stefan.fischer\}@scch.at}
}
\maketitle

\begin{abstract}
Metamorphic Testing (MT) is a testing technique that can effectively alleviate the oracle problem. MT uses Metamorphic Relations (MRs) to determine if a test case passes or fails. MRs specify how the outputs should vary in response to specific input changes when executing the System Under Test (SUT). If a particular MR is violated for at least one test input (and its change), there is a high probability that the SUT has a fault. On the other hand, if a particular MR is not violated, it does not guarantee that the SUT is fault free. However, deciding if the MR is being violated due to a bug or because the MR does not hold/fit for particular conditions generated by specific inputs remains a manual task and unexplored. In this paper, we develop a method for refining MRs to offer hints as to whether a violation results from a bug or arises from the MR not being matched to certain test data under specific circumstances. In our initial proof-of-concept, we derive the relevant information from rules using the Association Rule Mining (ARM) technique. In our initial proof-of-concept, we validate our method on a toy example and discuss the lessons learned from our experiments. Our proof-of-concept demonstrates that our method is applicable and that we can provide suggestions that help strengthen the test suite for regression testing purposes.

\end{abstract}


\section{Introduction}
\label{sec:Introduction}

Software testing is a crucial stage in the software development life cycle as it ensures the software's proper operation and quality. One of the most significant challenges in software testing is the test oracle problem. A test oracle determines the System Under Test's (SUT) output for a given input. The test oracle problem occurs when the SUT lacks an oracle or when creating one to verify the computed outputs is practically impossible \cite{DuqueTorres2020UsingRM}. \textit{Metamorphic Testing} (MT) is a software testing approach proposed by \citeauthor{chen2020metamorphic}~\cite{chen2020metamorphic} to alleviate the test oracle problem. 

In contrast to traditional testing techniques, MT examines the relations between input-output pairs of consecutive SUT executions rather than the individual outputs. The relations between SUT inputs and outputs in MT are known as \textit{Metamorphic Relations} (MRs). MRs specify how the outputs should vary in response to specific input changes \cite{9825836}. When an MR is violated for at least one test input and its change, there is a strong likelihood that the SUT has a fault. However, the absence of violation does not ensure that the SUT is fault-free. As a result, the suitability of the MRs employed significantly impacts the effectiveness of MT \cite{9825836}.

In current practice, the identification and selection of MRs are made manually, requiring a deep understanding of the SUT and its problem domain. The requirement for domain knowledge makes automatic MR identification challenging \cite{9825836,9825846,}. Another critical challenge is the need to distinguish automatically whether a particular MR violation is due to a fault in the SUT or because the MR does not satisfy or fully fit a specific statement/method/function of the SUT for certain test data. In current practice, interpreting an MR violation is an entirely manual effort. It is important to highlight that the cost, in terms of time and resources, of the MT approach is related to the amount of MRs used \cite{prioritization1, 8539189}. Thus, as the number of MRs grows, the number of test cases may grow exponentially. As a result, the execution time and the time needed for manual inspection of MR violations will also increase. 

Some approaches indirectly reduce the manual effort required to interpret the meaning of an MR violation. For instance, \citeauthor{6605921}~\cite{6605921} provides quantitative suggestions/guidance for developing automated means of selecting/prioritising MR for cost-effective MT. \citeauthor{prioritization1}~\cite{8539189, prioritization1} proposed two MR prioritisation approaches to improve MT's efficiency and effectiveness. These approaches use (i) fault detection information and (ii) statement/branch coverage information to prioritise MRs. \citeauthor{8919101}~\cite{8919101} suggested strategies to clean MRs by deleting duplicate or redundant MRs.  
These approaches offer indirect help since by prioritising or reducing the set of MRs, the number of test cases will be reduced as well. Thus,  the manual effort of inspection through the violated MRs is less.

Motivated by the above, we ask ourselves the following research question: \textit{How can MRs be refined based on test data?}. To answer this question, we developed a method for refining MRs that suggest whether a detected MR violation results from a fault in the SUT or arises from the fact that the MR does not apply to the used test data. Our method assumes that a predefined set of MRs is provided and uses the concepts of fuzz testing, passive testing, and rule mining.

First, our method uses a fuzzer to feed random data to the SUT. Second, it performs the necessary input transformations following the indications of the MRs. Third, similar to passive tests, logs are produced with information related to inputs, outputs, and whether or not MRs are violated. Those logs are used to feed a mining algorithm. Our method employs association rule mining (ARM). In our context, the purpose of ARM is to extract interesting relationships between the inputs and whether or not the MR is violated. ARM is an unsupervised machine learning (ML) method \cite{ZHOU2007737}. ARM algorithms attempt to find relationships or associations between categorical variables in large transactional datasets \cite{7079081}. We were particularly interested in understanding whether the information provided by the resulting model helps in deciding whether there is a fault or whether the MR does not fully fit the specific method/function/statement when the violation occurs. 
% We address our objective by answering the following research questions:

% \textbf{RQ$_{1}$:} \textit{How effective is the rule mining approach for refining MRs?}. This research question investigates the extent to which ARM is able to detect if there is a fault or if the MR does not fully conform to the specific method/function/statement when the violation occurs.

% \textbf{RQ$_{2}$:} \textit{How effective is our approach for regression testing purpose?} This research question explores to what extent the information contained in association rules helps to create/improve and strengthen the test suite for regression testing purposes.

In our initial proof-of-concept, we validate our method on a toy example and discuss the lessons learned from our experiments. Our proof-of-concept demonstrates that our method is applicable and can provide suggestions that help strengthen the test suite for regression testing purposes. We published the replication package to facilitate future research.

The rest of the paper is structured as follows. \Cref{sec:Background} presents the main concepts used in our research. In \Cref{sec:Proposed methodology}, we describe the proposed method. In \Cref{sec:Results and Discussion}, we present our results and discuss threats to validity. \Cref{sec:Relatedwork} presents the related work. Finally, we conclude the paper in \Cref{sec:Conclusion and future work}.

\section{Background}
\label{sec:Background}

This section presents the key concepts used in our research. \Cref{subsec:Metamorphic Testing} introduces the MT approach. \Cref{subsec:TDG} provides a brief description of test data generation techniques, and \Cref{subsec:ARM} gives a brief introduction to ARM.
% and in \Cref{subsec:Fuzzing-Testing} and \Cref{subsec:Passive-testing} we briefly describe fuzzing testing and passive testing.

\subsection{Metamorphic Testing}
\label{subsec:Metamorphic Testing}
\begin{figure}[tp!]
	\centering
	\includegraphics[width=\linewidth]{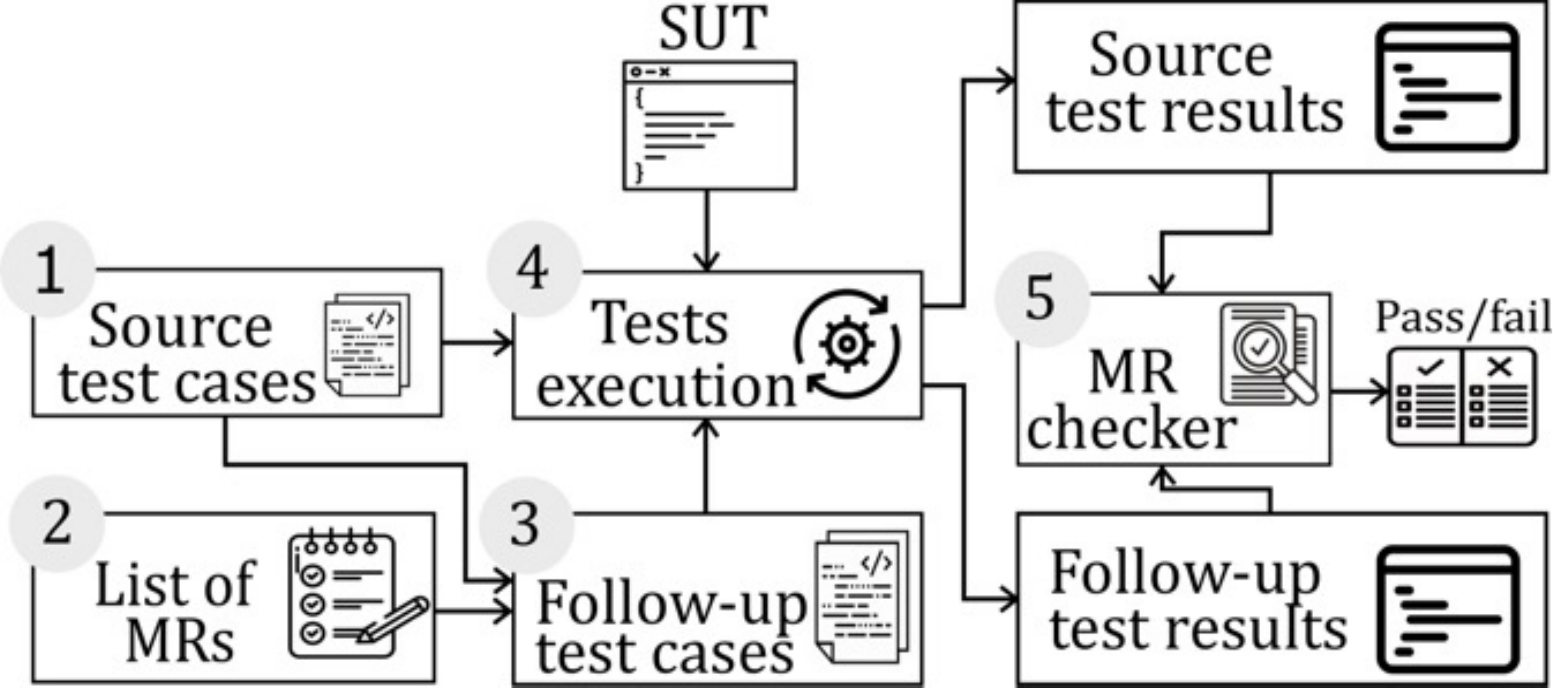}
	\caption{MT basic workflow}
	\label{fig:MT}
% 	\vspace*{-3ex}
\end{figure}

MT is a software testing approach that alleviates the test oracle problem. MT aims to exploit the internal properties of a SUT to either check its expected outputs or generate new test cases. \Cref{fig:MT} shows the MT basic workflow. Overall, MT works by checking the relation between the inputs and outputs of multiple executions of the SUT. Such relations are called MRs. MRs specifies how the outputs should change according to specific variations made to the input. Overall, MT follows four major steps:
\begin{enumerate}
    \item  Create a set of initial tests or source test cases.
    \item Identify an appropriate list of MRs that the SUT should satisfy.
    \item Create follow-up test cases by applying the input transformations required by the identified MRs in Step 2 to each source test case.
    \item Execute the corresponding initial and follow-up test case pairs.
    \item Check if the source tests and follow-up tests output change matches the change predicted by the MR.
\end{enumerate}

The final step needs further interpretation of the MT workflow output based on the Non-violation of MRs. When a Non-violation is presented, it is not guaranteed that the SUT is implemented correctly. However, if an MR is violated for specific test cases,  it must be a fault in the SUT, assuming the MR is defined correctly.  

\subsection{Test Data Generation}
\label{subsec:TDG}

In software testing, test data corresponds to the input data used during test execution. Test data is used for positive testing, verifying that functions of the SUT generate anticipated outputs for given inputs, and negative testing, examining the SUT's capacity to handle atypical, extraordinary, or unexpected inputs [ref]. Inadequately constructed testing data could only cover some potential test cases, which would be detrimental to the software's quality. 

Our method does not attempt to add something new in test data generation techniques. Instead, we take advantage of Fuzz Testing for generating test data. Fuzz Testing, often known as ``fuzzing," is a software testing approach involving injecting erroneous or random data, or ``FUZZ," into software systems to find coding errors and security flaws. Fuzz testing involves injecting data using automated or somewhat automated methods and evaluating the system for various exceptions, such as system failure or malfunction of built-in code, etc \cite{FuzzingSurvey}. 

Overall, fuzzing consists of three components, i.e., \textit{input generator}, \textit{executor}, and \textit{defect monitor} \cite{FuzzingSurvey}. The input generator provides the executor with several inputs, and the executor runs target programs on the inputs. Then, fuzzing monitors the execution to check if it discovers new execution states or defects (e.g., crashes). Fuzzing can be divided into Generation-based and Mutation-based fuzzing. Mutation-based fuzzing alters existing data samples to create new test data. Generation-Based fuzzing defines new data based on the system's input or the target SUT function. It starts generating input from scratch based on the specification.

\begin{figure*}[ht!]
	\centering
 	\includegraphics[width=\linewidth]{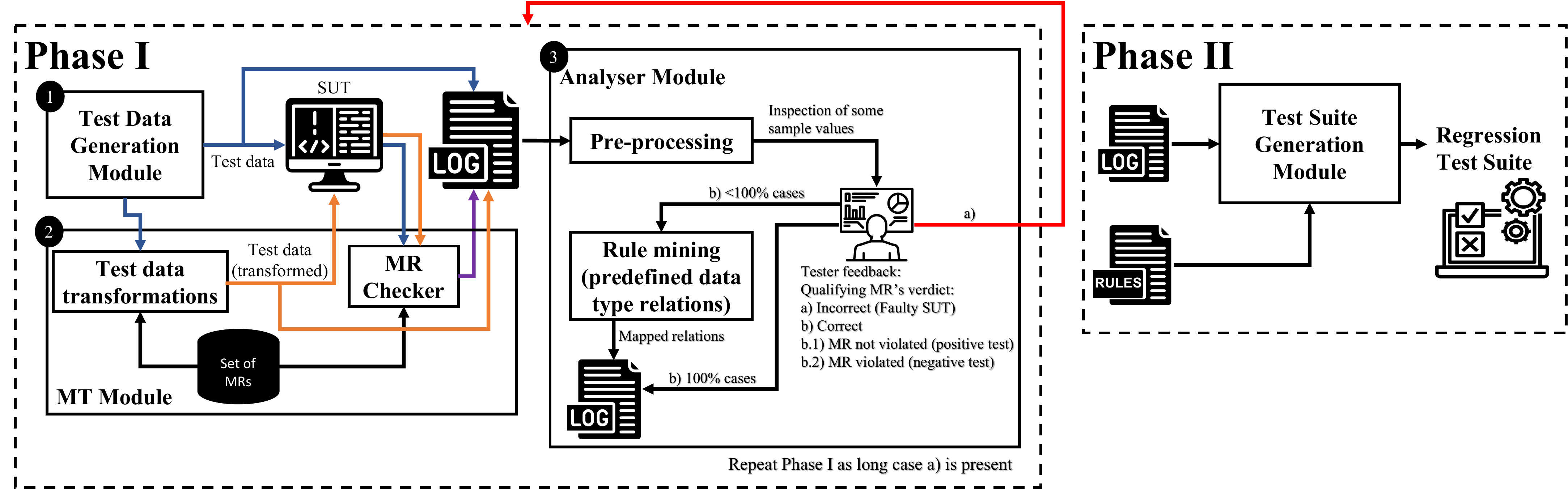}
	\caption{Overview of the method for refining MRs based on rule mining}
	\label{fig:method}
% 	\vspace*{-5ex}
\end{figure*}

\subsection{Association Rule Mining}
\label{subsec:ARM}
ARM  is a rule-based unsupervised ML method that allows discovery relations between variables or items in large databases. ARM has been used in other fields, such as business analysis, medical diagnosis, and census data, to find out patterns previously unknown \cite{7079081}. The ARM process consists of at least two major steps: finding all the frequent itemsets that satisfy minimum support thresholds and generating strong association rules from the frequently derived itemsets by applying a minimum confidence threshold. 

A large variety of ARM algorithms exist. \cite{10.5555/645920.672836}. In our experiments, we use the Apriori algorithm from Python3 Efficient-Apriori library \cite{apriori}. It is well known that the Apriori algorithm is exhaustive; it finds all the rules with specified support and confidence. In addition, ARM doesn't require labelled data and is, thus, fully unsupervised. Below we define important terminology regarding ARM:

\textbf{\textit{Itemset:}} Let \textit{X$_i$} be items, then \textit{I} =\{X$_1$,…, X$_k$\} is an Itemset of \textit{k} different items, with k $>$ 1.
	
\textbf{\textit{Association rule:}} Consider a dataset $D$, having $m$ different types of items and $n$ transactions defined by the itemsets constructed from the items. An association rule exposes the relationships between the elements of the itemsets in the set of $n$ transactions.
		
\textbf{\textit{Support:}} The support of an association rule involving itemsets X and Y is the percentage of transactions in dataset D that contain itemsets X and Y. The support of an association rule $X \rightarrow Y$: 

	$support(X \rightarrow Y) = support(X \cup Y)=P(X \cup Y)$
	
\textbf{\textit{Confidence: }}The confidence is the percentage of transactions in the dataset D with itemset X that also contains the itemset Y. The confidence is calculated using the conditional probability, which is further expressed in terms of itemset support: $confidence(X \rightarrow Y )= P(Y | X)=support(X \cup Y)/support(X)$
	
\textbf{\textit{Lift: }}Lift is used to measure the frequency of the occurrence of $X$ and $Y$ together if both are statistically independent of each other. The lift of rule $(X \rightarrow Y)$ is defined as $lift(X \rightarrow Y)=confidence(X \rightarrow Y)/support(Y)$. 

A lift value of 1 indicates that $X$ and $Y$ appear as frequently together as they appear individually under the assumption of conditional independence.

\section{Method}
\label{sec:Proposed methodology}

\Cref{fig:method} presents an overview of the method for refining MRs based on rule mining. In general, the proposed method comprises two phases. Phase I is in charge of identifying the degree of applicability of the set of MRs selected. It is important to highlight that our method does not cover the selection of appropriate MRs. We assume that there is already a predefined set of several MRs. Phase II uses the refined MRs to create or improve the test suite for future SUT versions. Thus, the output of this phase could be seen as a regression test suite. Each phase is thoroughly described below:

\subsection{\textbf{Phase I}} 
\label{subsec:phaseI}

Phase I comprises three modules: Test Data Generation Module (TDG Module), Metamorphic Test Module (MT Module) and Analyser Module. Overall, the TDG Module produces the test data that will feed the SUT and the MT Module. In the MT Module, the test data is transformed based on the indications of each MR; then, such transformed data is executed against the SUT. Both SUT outputs, the output produced with test data and the transformed test data, are executed against SUT, are checked against the MRs in the MR Checker. Then, the test data and the results of the MR Checker Module are organised and stored in a Log file. The Log file is used in the Analyser Module, where processed and refined MRs based on rules are extracted. These modules, their internal settings and their activities are detailed below:

\subsubsection{\textbf{TDG Module}}
\label{subsubsec:TestDataGenerationModule}
This module generates the test data, which will feed the SUT. It is important to note that our method does not try to add something new in the field of test data generation techniques. As we explained in \Cref{subsec:TDG}, our method uses fuzzing testing to generate test data. Overall, the basic fuzzing workflow involves three basic components, \textit{input generator}, \textit{executor}, and \textit{defect monitoring}. There are open-source tools that can be used in this module. However, it is necessary to consider the SUT's application domain. For instance, fuzzers such as SPIKE proxy, Peach Fuzzer, and OWASP WSFuzzer are highly recommended for security purposes in web systems. Also, the programming language must be considered when selecting the fuzzer. For instance, OSS-Fuzz, Google's open-source fuzzing platform, supports Java and Python language for finding security vulnerabilities, stability issues, and functional bugs. Regardless of the Fuzzer used, this module's most important is storing the generated test data.

\subsubsection{\textbf{MT Module}}
\label{subsubsec:MTModule}
Overall, this module is responsible for performing the test data transformation based on the changes in the inputs specified by the MRs, executing the transformed test data, and checking if the test data and the transformed test data outputs match the change predicted by the MRs. 
This module has three main activities, \textit{Set of MRs}, \textit{Test data transformations}, and \textit{MR Checker}.

\begin{itemize}
    \item \textbf{\textit{Set of MRs:}} It is important to note that our approach does not involve the initial selection of the MRs. Our approach assumes the prior existence of a predefined set of MRs.

    \item \textit{\textbf{Test data transformation:}} This activity is in charge of transforming the test data according to the change specified by each MR. To the best of our knowledge, no tool performs this activity automatically, \textit{i.e.,} the translation of input change described by the MR and its meaning  into code. Thus, this is considered to be a manual task.
    
    \item \textit{\textbf{MR Checker:}} This activity is responsible for checking that both outputs, test data and transformed test data match the change predicted by the corresponding MR. 
\end{itemize}

Once the test data is generated, transformed, and compared by the MR Checker, \textit{i.e.,} the verdict of the MR Checker is ready, a Log file is produced with the following information: execution ID, test data, function call, and the MR Checker verdict per MR. 

\subsubsection{\textbf{Analyser Module}}
\label{subsubsec:AnalyserModule}

The Analyzer Module is in charge of discovering interesting relations between test data and whether or not a certain MR has been violated. This module has three main activities, \textit{Pre-processing}, \textit{Tester feedback}, and \textit{Rule mining}. Below we describe each activity in detail as well as their internal process:

\begin{itemize}
    \item \textbf{\textit{Pre-processing:}} This activity is responsible for ensuring that the data is correct, consistent and usable. Also, it shows the tester an initial summary of the percentage of violations and not violations per each MR and function call. This activity has three main functions: \textit{data quality}, \textit{clean}, and \textit{summary report}. The \textit{data quality} function is responsible for checking that Log has no missing data, as well as removing the rows that are not needed or inconsistent rows. The \textit{clean} removes duplicate entries. The \textit{Summary report} is based on the percentage of violations and no-violation per MRs. Also, it provides the tester with the ability to inspect some random sample values and atypical values. This is done to increase confidence in the suitability of the MRs.  
    
    \item \textbf{\textit{Tester feedback:}} This activity is in charge of qualifying MR's verdict based on the summary report provided by the \textit{Pre-processing} activity. Here the tester needs to perform two checks when there is an incorrect and correct  behaviour of the MR Checker output: 

    \begin{itemize}
        \item [a)] \textit{Incorrect behaviour}: The tester evaluates whether there is an obvious fault in the SUT. Phase I should be repeated as long the fault is present.
        \item [b)] \textit{Correct behaviour}: In the correct behaviour there are two possibilities.

        \begin{itemize}
            \item [b.1)] \textit{MR not violated} which represents a positive test.
            \item [b.2)] \textit{MR violated} which represents a negative test.
        \end{itemize}

    \end{itemize}
    If specific MR is violated 100\% of the time, we assume that the MR does not apply to the tested function. On the other hand, if it is not violated 100\% of the time, we assume that the MR matches the tested function. In both cases, the tester can directly decide whether to include them in the Rule file. Those 100\% violations could be used as negative tests and the 100\% of no violations as positive tests.
    We also look at the atypical values; for instance, if the MR was violated or not violated only 10\% of the time.

    \item \textit{\textbf{Rule mining (predefined data type relations):}}  This activity is responsible for generating the rule set by discovering interesting relationships between the test data and whether or not a particular MR has been violated. This activity has three internal steps:
    \begin{itemize}
        \item[1)] \textit{Encoding:} This step is in charge of preparing the data according to the requirements of the rule mining algorithm. For example, Apriori \cite{BHANDARI2015644}, which is the algorithm used in this paper, works only with categorical features. Thus, this component categorises and generalises the numerical inputs into string representations. 

        \item[2)] \textit{Rule generation:} This step is responsible for generating the set of rules using the Apriori ARM algorithm.

        \item[3)]\textit{Data type relation:} This step is in charge of generalising the data based on its data type relation. This data type is predefined. For example, the test data for SUT can be generalised using partial order theory. The partial order defines a notion of comparison between at least two elements, \textit{e.g.,} input$_a$ and input$_b$. The two elements input$_a$ and input$_b$ can be in any of four mutually exclusive relationships to each other: 
        \begin{description}
            \item input$_a$ $<$ input$_b$
            \item input$_a$ = input$_b$
            \item input$_a$ $>$ input$_b$
            \item input$_a$ and input$_b$ are incomparable
        \end{description}
        The latter relationship is not present in our data.
    \end{itemize}
\end{itemize}

\subsection{\textit{Phase II}}
Phase II is in charge of generating the test suite for regression testing purposes. Overall, this phase takes the Log, which has the test data generated and takes the Log with the set of rules for generating the final test suite.

\section{Results and Discussion}
\label{sec:Results and Discussion}
This section presents and discusses the preliminary results of our approach. Let's consider the \Cref{alg:SUT} as the SUT. \Cref{alg:SUT} is a program that computes three basic arithmetic operations, addition, subtraction, and multiplication between two integers. The full set of data generated during our experiments, as well as all scripts, can be found in our GitHub repo\footnote{\href{https://github.com/aduquet/VST2023-BugORNOTbug}{https://github.com/aduquet/VST2023-BugORNOTbug}}.

\begin{algorithm}[!ht]
	\caption{Calculator()}\label{alg:SUT}
	\begin{algorithmic}[1]
		\Procedure{AddSub}{int a, int b}                                \Function{add}{a, b}
		\State \textbf{return} a + b
		\EndFunction

        \Function{sub}{a, b}
		\State \textbf{return} a - b
		\EndFunction
		
		\Function{mul}{a, b}
		\State \textbf{return} a * b
		\EndFunction
		
		\EndProcedure
	\end{algorithmic}
\end{algorithm}

\subsection{\textbf{Phase I}} 
\subsubsection{\textbf{Test Data Generation Module}}
\label{subsubsec:TestDataGenerationModule_result}
\begin{figure}[ht!]
	\centering
 	\includegraphics[width=0.9\linewidth, trim=6mm 0mm 14mm 8.4mm]{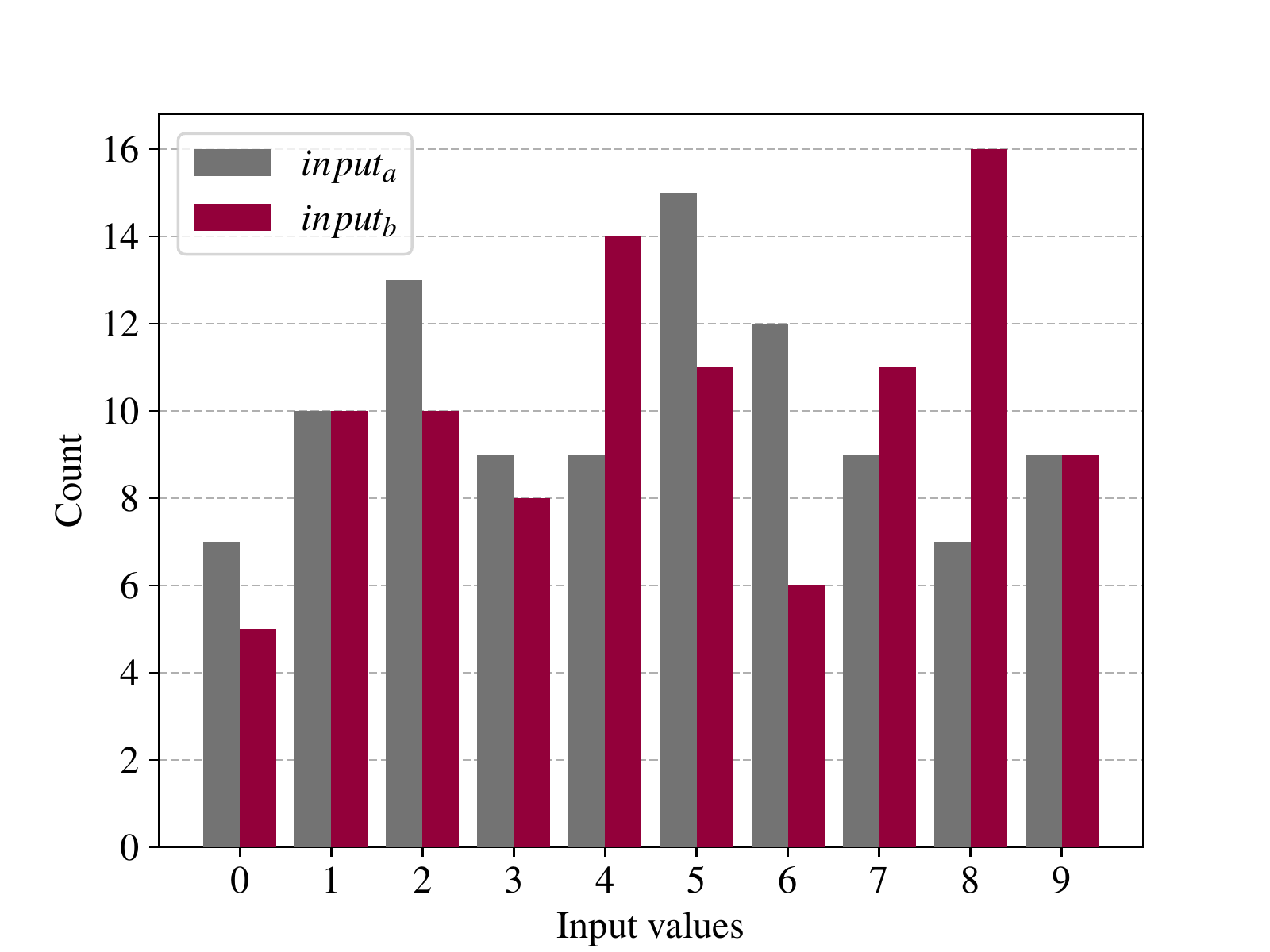}
	\caption{Distribution of data generated for input$_a$ and input$_b$}
 
	\label{fig:inputs}
    \end{figure}

In this module, we create a fuzzer based on a random number generator. For this, we use the NumPy random function in Python.
A total of 100 random numbers, elements of \{0, 1, 2, ..., 9\}, were generated for input$_a$ and input$_b$, following a uniform distribution. Since, for some MRs, a constant is needed, it was generated only once and reused every time it was needed. \Cref{fig:inputs} shows the histogram of the generated test data, \textit{i.e.,} input$_a$ and input$_b$. 

 %    \begin{figure}[ht!]
	% \centering
 % 	\includegraphics[width=\linewidth]{Figs/input_a.png}
	% \caption{Distribution of data generated for input$_a$ }
	% \label{fig:input_a}
 %    \end{figure}

 %    \begin{figure}[ht!]
	% \centering
 % 	\includegraphics[width=\linewidth]{Figs/input_b.png}
	% \caption{Distribution of data generated for input$_b$}
	% \label{fig:input_b}
 %    \end{figure}
    
\subsubsection{\textbf{MT Module}}
\label{subsec:MTModule_results}
\begin{itemize}
    \item \textit{\textbf{Set of MRs:}} \Cref{tbl:MRs_motivationexample} describes the set of MRs for \Cref{alg:SUT}. For our SUT, we take advantage of the generic rules of arithmetic to build a set of four MRs that may apply to the SUT. MR$_2$, MR$_3$, and MR$_4$ need a constant $k$. That constant was randomly generated only once and reused each time it was needed. 
    
\begingroup
\setlength{\tabcolsep}{6pt} % Default value: 6pt
\renewcommand{\arraystretch}{1} % Default value: 1
\begin{table}[ht!]
\centering
\caption{Set of MRs for \Cref{alg:SUT}}
{
	\label{tbl:MRs_motivationexample}
	\resizebox{\linewidth}{!} {
	\begin{tabular}{l|l|l}
		\toprule
		\textbf{MR} & \textbf{Change in the input} & \textbf{Expected output}\\
		\toprule
	    \textbf{MR$_1$} & Permute the inputs          & Remain equal \\
	    \textbf{MR$_2$} & Multiply by a positive constant $k$ $> 1$               & Increase \\
	    \textbf{MR$_3$} & Adding a positive constant $k$ to each operand               & Remain equal\\
		\textbf{MR$_4$} & Subtracting a positive constant $k$ from each operand & Remain equal \\
		\bottomrule
	\end{tabular}}}
\end{table}
\endgroup

    \item \textbf{\textit{Test data transformation}} \Cref{tbl:TestDataTransformed} shows how the inputs are transformed following the MR indications. 
    \item \textit{\textbf{MR Checker:}} \Cref{tbl:MRs_Checker} shows how the MR Checker checks the outputs following the expected output predicted by the MRs and gets the verdict, \textit{i.e.,} not violated or violated for our SUT.

\begingroup
\setlength{\tabcolsep}{6pt} % Default value: 6pt
\renewcommand{\arraystretch}{1.2} % Default value: 1
\begin{table*}[ht!]
\centering
\caption{Test data transformations of the test data according with the MRs described in \Cref{tbl:MRs_motivationexample} for \Cref{alg:SUT}}
{
	\label{tbl:TestDataTransformed}
	\resizebox{\linewidth}{!} {
	\begin{tabular}{l|l|l|l}
		\toprule
		\textbf{MR} & \textbf{Input change} &\textbf{Test data (input$_a$, input$_b$)} & \textbf{transformed test data (T-input$_a$, T-input$_b$)}   \\
            % \textbf{MR} & \textbf{MR input change} &\textbf{Test data} & \textbf{transformed test data} \\
		\toprule
	    \textbf{MR$_1$} & Permute the inputs & \multirow{4}{*}{$^\dagger$input$_a$ = a, input$_b$ = b} & T-input$_a$ = b, T-input$_b$ = a \\
	    \textbf{MR$_2$} & Multiply by a positive constant $k$  $> 1$ & & T-input$_a$ = a * $k$ , T-input$_b$ = b * $k$  \\
	    \textbf{MR$_3$} & Adding a positive constant $k$ to each operand & & T-input$_a$ = a + $k$ , T-input$_b$ = b + $k$  \\
		\textbf{MR$_4$} & Subtracting a positive constant $k$ from each operand & & T-input$_a$ = a - $k$ , T-input$_b$ = b - $k$  \\
		\bottomrule
  \multicolumn{4}{l}{{$^\dagger$ a, b are elements of \{0, 1, 2, ..., 9\}}}\\
	\end{tabular}}}
\end{table*}
\endgroup

\begingroup
\setlength{\tabcolsep}{6pt} % Default value: 6pt
\renewcommand{\arraystretch}{1.2} % Default value: 1
\begin{table*}[ht!]
\centering
\caption{MR Checker verdict example according with the expected outputs predicted by the MRs described in \Cref{tbl:MRs_motivationexample} for \Cref{alg:SUT}}
{
	\label{tbl:MRs_Checker}
	\resizebox{\linewidth}{!} {
	\begin{tabular}{l|l|l|l|l|l}
		\toprule
		\textbf{MR} & \textbf{Expected output} &\textbf{Test data} & \textbf{transformed test data} & \textbf{MR Checker}& \textbf{Verdict}\\
            % \textbf{MR} & \textbf{MR input change} &\textbf{Test data} & \textbf{transformed test data} \\
		\toprule
	    \textbf{MR$_1$} & Remain equal & \multirow{2}{*}{$^\dagger$input$_a$ = a} & T-input$_a$ = b, T-input$_b$ = a & SomeFunc(input$_a$,input$_b$) == SomeFunc(T-input$_a$, T-input$_b$) & \multirow{2}{*}{\textbf{True:} No-Violated}\\
	    \textbf{MR$_2$} & Increase  &  & T-input$_a$ = a * $k$ , T-input$_b$ = b * $k$ &
        SomeFunc(input$_a$,input$_b$) $<$ SomeFunc(T-input$_a$, T-input$_b$)\\
	    \textbf{MR$_3$} & Remain equal & \multirow{2}{*}{$^\dagger$input$_b$ = b}& T-input$_a$ = a + $k$ , T-input$_b$ = b + $k$ & 
        SomeFunc(input$_a$,input$_b$) $==$ SomeFunc(T-input$_a$, T-input$_b$) & \multirow{2}{*}{\textbf{False:} Violated} \\
		\textbf{MR$_4$} & Remain equal & & T-input$_a$ = a - $k$ , T-input$_b$ = b - $k$ &
        SomeFunc(input$_a$,input$_b$) $==$ SomeFunc(T-input$_a$, T-input$_b$) \\
		\bottomrule
  \multicolumn{4}{l}{{$^\dagger$ a, b are elements of \{0, 1, 2, ..., 9\}}}\\
  \multicolumn{4}{l}{SomeFunc refers to a function call of the class, i.e., ADD, SUB or MUL in \Cref{alg:SUT}}
	\end{tabular}}}
\end{table*}
\endgroup
    
\begingroup
\setlength{\tabcolsep}{6pt} % Default value: 6pt
\renewcommand{\arraystretch}{1} % Default value: 1
\begin{table}[ht!]
\centering
\caption{MR Checker output example}
{
	\label{tbl:MRs_checker_log}
	\resizebox{\linewidth}{!} {
	\begin{tabular}{l|l|l|c|c|c|c}
		\toprule
		\textbf{id} & \textbf{(input$_a$,input$_b$)} & \textbf{Func Call} & \textbf{MR$_1$} & \textbf{MR$_2$}&\textbf{MR$_3$} &\textbf{MR$_4$}\\
		\toprule
	    0 & (0,0) & add & \checkmark & \xmark & \xmark & \xmark \\
	    1 & (1,1) & sub & \checkmark & \xmark & \xmark & \xmark\\
            2 & (2,1) & mul & \checkmark & \xmark & \xmark & \xmark\\
	    ... & ... & ... & ... & ... & ... & ...\\
		299 & (9,9) & add & \checkmark & \checkmark & \xmark & \xmark\\
		\bottomrule
   \multicolumn{4}{l}{{\checkmark : Not violate, \xmark: Violated}}\\
	\end{tabular}}}
\end{table}
\endgroup
\end{itemize}

\subsubsection{\textbf{Analyser Module}}
\label{subsubsec:AnalyserModule_result}

\begin{itemize}
    \item \textit{\textbf{Pre-procesing:}} \Cref{fig:SummaryReportFeedback} shows an example of the summary report for \Cref{alg:SUT}, \textit{i.e.,} not violated or violated, with a controlled input space.

    \item \textit{\textbf{Tester feedback:}} 
    \Cref{fig:SummaryReportFeedback} shows the percentage of MR not violated and violated per function call. From the first line, which belongs to the ADD function, we can see that for MR$_3$ and MR$_4$, there are 100\% violations. In this case, we assume that MR3 and MR4 do not apply to the SUM function. However, we have two options: include a test (negative test) for all test data or not include it. on the other hand, MR1 and MR2 present 100\% and 99\% of not violation respectively. After inspecting random samples to check, we can conclude that MR$_1$ and MR$_2$ completely match the ADD function. This means that MR1 and 99\% of MR$_2$ will be directly included in the set of rules to build positive tests in phase II.
    The 1\% violation is a particular case for MR$_2$ in the ADD function, which occurs whenever input$_a$ and input$_b$ are equal to 0. 

\begin{figure}[ht!]
\centering
    \includegraphics[width=0.9\linewidth]{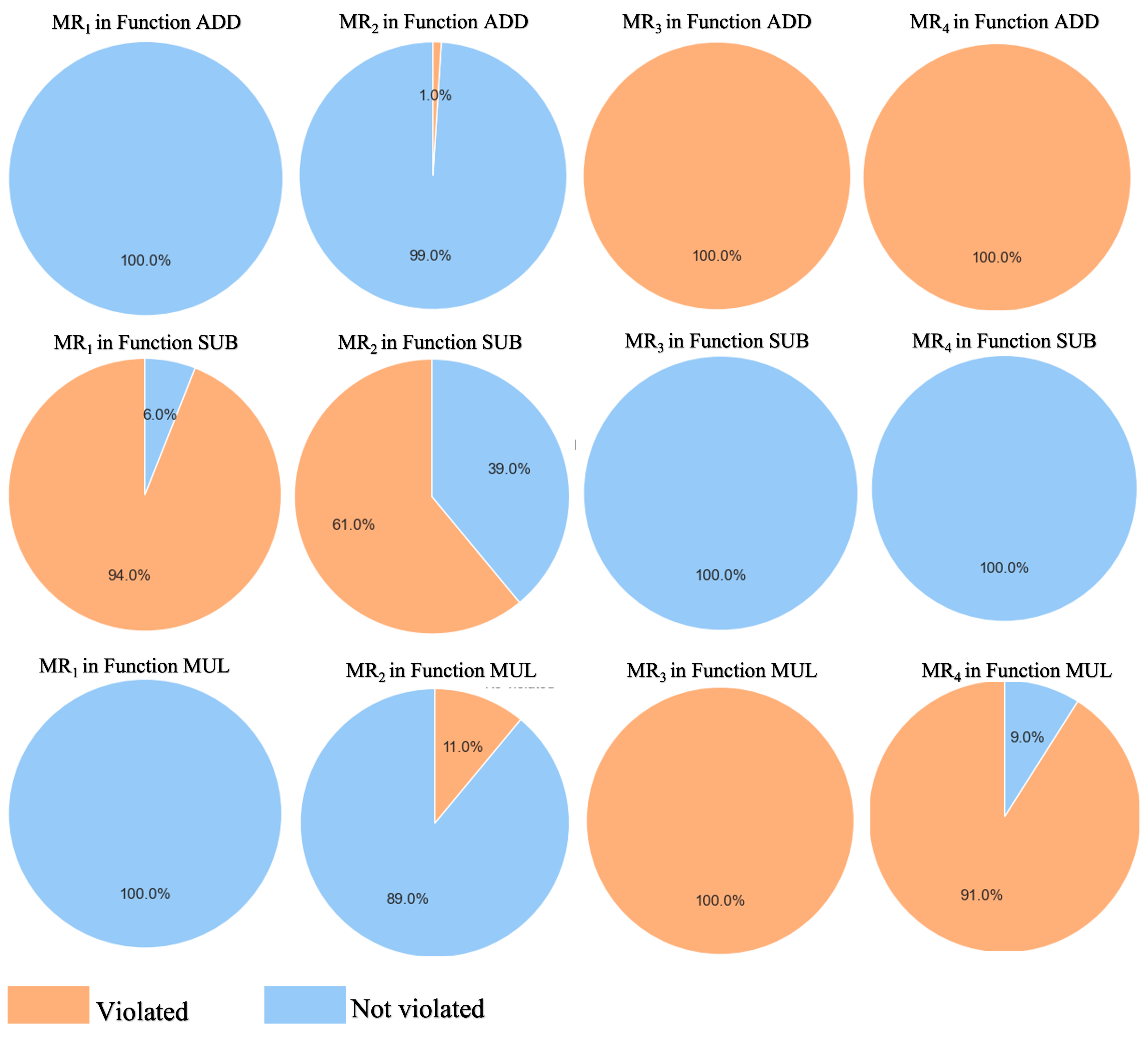}
    \caption{Summary report for the tester feedback}
    \label{fig:SummaryReportFeedback}
\end{figure}

    In the second line of pie charts in the \Cref{fig:SummaryReportFeedback}, which belongs to the SUB function, you can see the opposite behaviour of MR$_3$ and MR$_4$ compared to the ADD function. For these MR we have 100\% non-violations. This indicates that the MRs fully match the SUB function. Like MR$_1$ and MR$_2$ in the ADD function, MR$_3$ and MR$_4$ can go directly as a positive test for the SUB function in phase II. We can't say to much about MR$_1$ and MR$_2$ for SUB function. These MRs, \textit{i.e.,} MR$_1$ and MR$_2$, are the ones who we are interested to analyse with the \textit{rule mining}. 

    The results from MUL function are in the third line of the pie charts in the \Cref{fig:SummaryReportFeedback}. In this function, one can see that MR$_1$ and $MR_3$ behave the same as in the ADD function, which means that MUL also fully matches MR$_1$ and can be included directly in the final rules. It similarly happens with MR$_3$; it has the same behaviour as the ADD function, meaning we can treat it in the same way as MR$_3$ for ADD function. The 11\% of the violated cases in MUL function for MR$_2$ occurs whenever input$_a$ or input$_b$ are equal to 0. Here one could assume that we have other particular case in the test data. It indicates that we can use the this particular case to create a negative MR$_2$ test for MUL. It cannot be the same as in the ADD function since in the MUL function, these violations occur when either input$_a$ or input$_b$ is 0, and in ADD function, both inputs must be equal to zero. From the information in \Cref{fig:SummaryReportFeedback} one can assume the following:
    \begin{itemize}
        \item[a)] MR$_1$ applies to ADD and MUL function for all the test data.  
        \item[b)] MR$_2$ applies to the SUB function for almost all the test data, but when input$_a$ == input$_b$ == 0, MR$_2$ must be used as a negative test.
        \item[c)] MR$_3$ and MR$_4$ does not apply to the ADD function.
        \item[d)] MR$_3$ and MR$_4$ applies to the SUB function for all the test data. 
        \item[f)]MR$_3$ does not apply to the MUL function. 
    \end{itemize}
    
    \item \textit{\textbf{Rule mining:}} We apply the Apriori algorithm with minimal support and maximum confidence thresholds, \textit{i.e.}, 0.2 and 1. 
    We created the data type relation in the following way:
    \begin{description}
            \item input$_a$ $<$ input$_b$
            \item input$_a$ = input$_b$
            \item input$_a$ $>$ input$_b$
    \end{description}

% \begingroup
% \setlength{\tabcolsep}{6pt} % Default value: 6pt
% \renewcommand{\arraystretch}{1} % Default value: 1
% \begin{table}[ht!]
% \centering
% \caption{Set of rules for MR$_2$ in function SUB }
% {
% 	\label{tbl:RulesGeneratedforMR2SUB}
% 	\resizebox{\linewidth}{!} {
% 	\begin{tabular}{l|l|l|c|c}
% 		\toprule
% 		\textbf{RHS} & \textbf{LHS} &\textit{\textbf{conf}} & \textit{\textbf{sup}}& \textbf{\textit{lift}}\\
% 		\toprule
% 	    ('a $<$ b', SUB) & Violated  &1.0 &0.46&1.639 \\
% 	    ('a $>$ b', SUB) & Not violated & 1.0&0.32&2.564\\
% 	    ('a $==$ b', SUB)& Violated & 1.0 & 0.06 &1.639 \\
% 		\bottomrule
%    \multicolumn{4}{l}{{\textbf{RHS: }Right Hand Side, \textbf{LHS: }Left Hand Side}}\\
% 	\end{tabular}}}
% \end{table}
% \endgroup

\Cref{tbl:RulesGeneratedforFinal} shows the final set of detected rules. The first three rules are the ones that were directly inferred from the Tester Feedback. The other three rules are the output of the ARM. \Cref{tbl:RulesGeneratedforFinal} shows that the relation a$<$b has the largest support. This is due to the amount of samples that fall in that range. The lower support belongs to the relation a $==$ b, meaning that not many samples fall into that range.

\begingroup
\setlength{\tabcolsep}{6pt} % Default value: 6pt
\renewcommand{\arraystretch}{1} % Default value: 1
\begin{table}[ht!]
\centering
\caption{Final set of rules}
{
	\label{tbl:RulesGeneratedforFinal}
	\resizebox{\linewidth}{!} {
	\begin{tabular}{l|l|l|c|c}
		\toprule
		\textbf{RHS} & \textbf{LHS} &\textit{\textbf{conf}} & \textit{\textbf{sup}}& \textbf{\textit{lift}}\\
		\toprule
  	    ('a $==$ b $==$ 0', ADD) & MR$_2$ = Violated   \\
            ('a $==$ b', SUB) & MR$_2$ = Not violated   \\
	    ('a $==$ 0' or 'b $==$ 0', MUL) & MR$2$ = Violated   \\

	    ('a $<$ b', SUB) & MR$_2$ = Violated  &1.0 &0.46&1.639 \\
	    ('a $>$ b', SUB) & MR$_2$ = Not violated & 1.0&0.32&2.564\\
	    ('a $==$ b', SUB)& MR$_2$ = Violated & 1.0 & 0.06 &1.639 \\
		\bottomrule
   \multicolumn{4}{l}{{\textbf{RHS: }Right Hand Side, \textbf{LHS: }Left Hand Side}}\\
	\end{tabular}}}
\end{table}
\endgroup

\end{itemize}

\subsection{Phase II}
\begin{figure}[ht!]
    \centering
    \includegraphics[width=\linewidth]{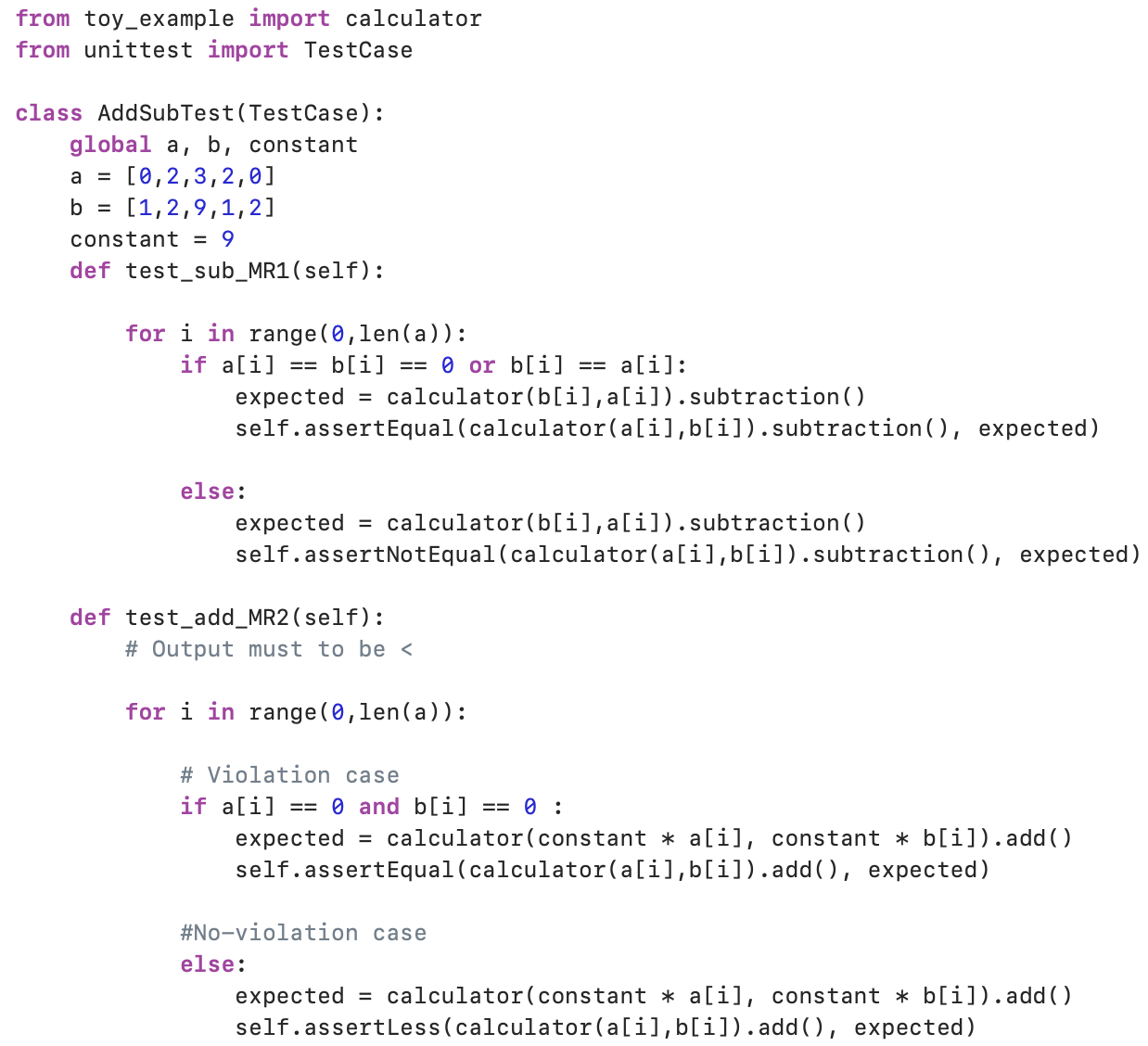}
    \caption{Example test suit using refined MRs}
    \label{fig:testSuite}
\end{figure}

With the information assumed in the feedback tester step and the \Cref{tbl:RulesGeneratedforFinal}, one can create a test suite for regression testing purposes. \Cref{fig:testSuite} shows an example of test code for the SUB function using MR$_1$ and the ADD function using MR$_2$. The full example can be found in our GitHub repository.

\subsection{Remarks on effectiveness}
In terms of MRs selection, our method cannot be worse than any existing method that manually create MRs. First, our method reuses the already created MRs. Second, our method can add more MRs by refining the existing ones with specific test data.
Regarding the inspection of the truthfulness of the test output, \textit{i.e.,} whether or not the MR is violated, it is important to note that using our approach, whatever the results of throwing data against the set of MRs, it will always be the same situation if the MRs have been manually created. In fact, it is part of the MT approach to inspect such results, \textit{i.e.,} whether or not the MR is violated.

The advantage of our approach is that it can provide hard facts on which MRs out of the set of MRs can and cannot be applied. In this regard, our approach reduces the time spent in selecting MRs, which is a manual and very time-consuming task. The downside of our approach is the time required for setup, the runtime of MRs against the test data, and manual inspection. However, the execution of the MRs with the test data can be automated, and the manual inspection is performed only once.

In terms of the effectiveness of our approach to finding defects, we face the same problem as any testing approach. It is well known that there is an open and old discussion about the dependency of test data and test cases. Any approach that uses MR can only find fault with them if the test data is chosen wisely. The proper selection of test data is planned to be explored in the future.

\subsection{Threats to validity}
In the context of our proof-of-concept validation, two types of threats to validity are most relevant: threats to internal and external validity.

\subsubsection{Internal validity}
\label{subsec:Internal_validity}
We used a well-known program as a proof-of-concept and the most common ARM algorithm to achieve internal validity. It is not fully clear in which situations the ARM is the best choice for our method. Future research in the direction of evaluating different ARM algorithms and feature selection needs to be done.

\subsubsection{External validity}
\label{subsec:External_validity}
 With regard to external validity, our study is rather limited since we only use one well-understood class in our experiments. Thus, the actual scope of the effectiveness of our proposed method is yet to be determined.

\subsubsection{Construct validity}
\label{subsec:construc_validity}
In this paper, we used the NumPy package, in particular its random function, to generate the test data. The usage of these third-party libraries represents potential threats to construct validity. To avoid this, we verified that the results produced by the random function are uniform by manually inspecting the generated distributions. 

\section{Related Work}
\label{sec:Relatedwork}
Since MT was introduced in 1998 by \citeauthor{chen2020metamorphic}~\cite{chen2020metamorphic}, MT has been demonstrated to be an effective technique for testing in a variety of application domains. Several studies have shown MT as a strong technique for testing the ``non-testable programs" where an oracle is unavailable or too difficult to implement \cite{segura2016survey,10.1145/3143561,Murphy2008PropertiesOM,8573811,Murphy2008PropertiesOM}. Also, MT has been demonstrated to be an effective technique for testing in a variety of application domains, \textit{e.g.,} autonomous driving \cite{zhang2018deeproad,zhou2019metamorphic}, cloud and networking systems \cite{canizares2020mt,9477667}, bioinformatic software \cite{10.1145/3193977.3193981, shahri2019metamorphic}, scientific software \cite{peng2021contextual}. However, the efficacy of MT heavily relies on the specific MRs employed and its interpretation of the meaning of MR violations.

As we mentioned before, some approaches indirectly reduce the manual effort required to interpret the meaning of an MR violation through prioritisation. For instance, \citeauthor{6605921}~\cite{6605921} provides quantitative suggestions/guidance for developing automated means of selecting/prioritising MR for cost-effective MT. \citeauthor{prioritization1}~\cite{8539189, prioritization1} proposed two MR prioritisation approaches to improve MT's efficiency and effectiveness. These approaches use (i) fault detection information and (ii) statement/branch coverage information to prioritise MRs. \citeauthor{8919101}~\cite{8919101} suggested strategies to clean MRs by deleting duplicate or redundant MRs.  
These approaches offer indirect help since by prioritising or reducing the set of MRs, the number of test cases will be reduced as well. Thus,  the manual effort of inspection through the violated MRs is less.

\section{Conclusion and future work}
\label{sec:Conclusion and future work}

We presented a new ARM-based method for refining MRs that suggest whether a detected MR violation results from a fault in the SUT or arises from the fact that the MR does not apply for the used test data. Our method assumes that a predefined set of MRs is provided and uses the concepts of fuzz testing and ARM. Our method consists of two phases. The purpose of Phase I is to evaluate the level of applicability of the chosen set of MRs. 

In phase I, there are three main modules: TDG Module, MT Module, and Analyser Module. The TDG Module generates the test data that will be sent to the MT Module and SUT. According to each MR's instructions, the test data is changed in the MT Module before being run against the SUT. In the MR Checker, the output from running test data and the transformed test data against SUT are compared to the changes predicted by the MRs. The test data and the MR Checker Module's results are then organised and saved in a Log file. The Analyser Module uses the Log file to process and improve MRs according to relations test data and whether or not the MR is violated.

Phase II is in charge of analysing the final set of rules and creating the new test suite. In our proof-of-concept, we used a toy example, a program that computes three basic arithmetic operations, addition, subtraction, and multiplication between two integers. We show step by step the execution of our method and its expected outputs from each module. 

An advantage of our method is that it can be applied not on the SUT with only integer inputs and outputs but on the class under test where we can have any inputs. Also, of mixed types of data. Thus, our method can be generalised for inputs of any type, not only for integers. It removes some limitations on the type of SUT that can be analysed. The weakness of our method is the need for manual feedback from the tester. However, compared with the manual effort already needed in the MT approach, we consider that the effort needed in our approach is less. We must manually translate the rules into assertions (test code) to generate the final test suite. For efficiency reasons, it would be better to have an automatic translation of the rules generated into test code. Unfortunately, there is no simple way to do this. 

Given the limitations of our study, more experiments have to be conducted to test our proposed method empirically. We are currently focusing on extending our experiments in three directions. First, we will add more MRs in the initial set of MRs to test the sensitivity of our method with regard to the filtering MRs that have no relation to the SUT. We will systematise this by using the relation between the input type and the MR. Second, we will apply our proposed method to more SUT. Third, we will do a mutation analysis to evaluate the effectiveness of our approach.

\section*{Acknowledgement}
This research was partly funded by the Estonian Center of Excellence in ICT research (EXCITE), the IT Academy Programme for ICT Research Development, the Austrian ministries BMVIT and BMDW, the Province of Upper Austria in the frame of the Software Competence Center Hagenberg (SCCH), and grant PRG1226 of the Estonian Research Council. 

\balance
\printbibliography

\end{document}